\documentclass[ALICE,manyauthors]{etc/cernphprep}
\usepackage{etc/cernphprep}
\usepackage{orcidlink}

\begin{document}

\begin{titlepage}
\PHyear{2023}       
\PHnumber{131}      
\PHdate{07 July}  

\title{Measurement of inclusive charged-particle jet production in pp and \pPb collisions at \fivenn}
\ShortTitle{Charged-particle jet production in pp and p--Pb collisions}

\Collaboration{ALICE Collaboration\thanks{See Appendix~\ref{app:collab} for the list of collaboration members}}
\ShortAuthor{ALICE Collaboration} 

\begin{abstract}

Measurements of inclusive charged-particle jet production in \pp and \pPb collisions at center-of-mass energy per nucleon--nucleon collision \fivenn and the corresponding nuclear modification factor $\RpPbjch$ are presented, using data collected with the ALICE detector at the LHC.
Jets are reconstructed in the central rapidity region $\abs{\etaj}<0.5$ from charged particles using the \akT algorithm with resolution parameters $R=0.2$, $0.3$, and $0.4$.
The $\pt$-differential inclusive production cross section of charged-particle jets, as well as the corresponding cross section ratios, are reported for \pp and \pPb collisions in the transverse momentum range $10<\pTjch<140$~\GeVc and $10<\pTjch<160$~\GeVc, respectively, together with the nuclear modification factor $\RpPbjch$ in the range $10<\pTjch<140$~\GeVc.
The analysis extends the $\pt$ range of the previously-reported charged-particle jet measurements by the ALICE Collaboration.
The nuclear modification factor is found to be consistent with one and independent of the jet resolution parameter with the improved precision of this study, indicating that the possible influence of cold nuclear matter effects on the production cross section of charged-particle jets in \pPb collisions at \fivenn is smaller than the current precision.
The obtained results are in agreement with other minimum bias jet measurements available for RHIC and LHC energies, and are well reproduced by the NLO perturbative QCD \Pwg calculations with parton shower provided by \Pyeight as well as by \Scape simulations.

\end{abstract}
\end{titlepage}

\setcounter{page}{2} 



\section{Introduction}%
\label{sec:intro}

In high-energy hadronic collisions, scattering processes at very large momentum transfer $Q^{2}$ between quarks and gluons of the colliding nucleons produce parton showers, which subsequently fragment into collimated sprays of hadrons called jets.
Studies of jet production in proton--proton (\pp) collisions allow one to test the fixed-order perturbative quantum chromodynamics (pQCD) calculations of the jet production in the TeV domain and tune the higher order effects in QCD-based Monte Carlo (MC) event generators~\cite{Dasgupta:2016bnd,Kang:2016mcy,Liu:2017pbb,Currie:2016bfm}.
Furthermore, such studies, especially at low $\pt$, constrain the non-perturbative contributions, such as the hadronization and underlying event effects, to the inclusive jet cross section.
In addition, measurements in \pp collisions also provide the baseline for similar measurements in proton--nucleus (pA) and nucleus--nucleus (AA) collisions.
Comparing the jet production between \pp and pA collisions allows for an assessment of the effects related to the presence of bound nucleons in the colliding system, denoted as cold nuclear matter (CNM) effects at the initial state of the collisions, which can be described by partonic rescattering~\cite{Vitev:2003xu} and by modification of parton distribution functions (PDFs)~\cite{Connors:2017ptx}.
The study of these CNM effects is interesting in its own right since it is necessary to decouple the CNM effects from those related to the creation of the quark--gluon plasma (QGP), which is a hot and dense color-deconfined QCD matter created in AA collisions~\cite{Busza:2018rrf,Jacak:2012dx}.

The production of the QGP in AA collisions is confirmed by many observations (see in~\cite{ALICE:2022wpn} and the references therein).
One of the QGP signatures is the so-called jet quenching phenomenon. It is manifested by the suppression of high-$\pt$ hadron and jet yields with respect to those in \pp collisions~\cite{GYULASSY1990432,Qin:2015srf}.
One microscopic picture of this phenomenon assumes that, while traversing through the QGP, the initial highly-energetic parton loses energy via medium-induced gluon radiations and elastic scatterings with constituents of the hot and dense medium.
A convenient observable to quantify these jet quenching effects is the nuclear modification factor, defined as the ratio of the jet (or final-state hadron) yield produced in AA or pA collisions to that in \pp collisions, scaled by the average number of nucleon--nucleon collisions $\avg{\Ncoll}$~\cite{Miller:2007ri}.
A deviation from one of this ratio at high $\pt$ indicates the presence of nuclear effects.

Initially, the pA system was thought to be too small to create a QGP.
However, recent measurements show evidence of collective behavior in high-multiplicity \pp and \pPb collisions at the LHC~\cite{ALICE:2014dwt,ALICE:2012eyl,ATLAS:2012cix,CMS:2015yux} and in light nucleus--Au collisions at RHIC~\cite{PHENIX:2014fnc,STAR:2014qsy}.
By contrast, jet quenching phenomena have not yet been seen in small collision systems.
The question of possible QGP formation in small collision systems remains open and calls for further, more precise jet quenching searches.
The production of charged-particle jets and the corresponding nuclear modification factor $\RpPbjch$ in \pPb collisions at center-of-mass energy per nucleon--nucleon collision \fivenn were previously reported by the ALICE Collaboration based on Run $1$ data~\cite{ALICE:2015umm}.
The scaled $\pt$-differential charged-particle jet production cross section from \pp collisions at \seven was adopted for calculating the $\RpPbjch$ in this study, and results show that the $\RpPbjch$ is consistent with one within uncertainties.
The measurement of full jet production in \pPb collisions at $\sNN = 5.02$~\TeV has been presented by the ATLAS~\cite{ATLAS:2014cpa} and CMS Collaborations~\cite{CMS:2016svx} at the LHC, and in \dAu collisions at $\sNN = 200$~\GeV by the PHENIX Collaboration~\cite{PHENIX:2015fgy} at RHIC.

In this article, we revisit previous ALICE analyses by measuring the charged-particle jet production in the larger Run $2$ datasets of \pp and \pPb collisions at \fivenn, exploiting the excellent tracking capabilities of ALICE~\cite{ALICE:2010tia,ALICE:2014sbx}.
These large data samples enable higher precision measurement of the charged-particle jet production over a broader $\pTjch$ interval compared to the previous one~\cite{ALICE:2015umm}, extending $\pTjch$ down to $10$~\GeVc and up to $140$~\GeVc.
Furthermore, in contrast to the results from Run $1$, the reported $\RpPbjch$ from Run $2$ utilizes a \pp reference measured at the same collision energy.
Therefore, when constructing the nuclear modification factor, there is no need to rely on an interpolation between collision energies; instead, a more direct comparison can be made at the same $\snn$.
This reduces the normalization uncertainty by a factor of about $2.7$ as compared to the previously published results.
Data from \PbPb collisions are also available at the same energy~\cite{ALICE:2023waz, ALICE:2019qyj} and thus the results from this \pPb analysis can provide a baseline for the \PbPb data.
These measurements can also be used to constrain the nuclear-modified parton distribution functions (nPDFs)~\cite{Ball:2018iqk,CMS:2018jpl,Armesto:2015lrg, Paukkunen:2014nqa} and the strong coupling constant $\alpha_{\rm s}$~\cite{Britzger:2017maj}.

This article reports the measurement of charged-particle jet production both in \pp and \pPb collisions at \fivenn and the corresponding nuclear modification factor for jet resolution parameters $R=0.2$, $0.3$, and $0.4$.
The inclusive jet cross section is used to evaluate ratios of jet yields obtained for different resolution parameters.
These ratios provide insight into the interplay between perturbative and non-perturbative effects on jet transverse momentum scales~\cite{Dasgupta:2007wa,CMS:2014nvq,CMS:2020caw,Dasgupta:2016bnd}.

The paper is organized as follows.
\CSec{sec:setup} describes the ALICE detector and the dataset.
Jet reconstruction approach, correction for detector and acceptance efficiency, and systematic uncertainty assessment are discussed in \CSec{sec:dana}.
\CSec{sec:res} presents the results and compares them to theoretical predictions and other experimental measurements.
The conclusion is given in \CSec{sec:con}.


\section{Experimental setup and datasets}%
\label{sec:setup}

The ALICE detector is a general-purpose heavy-ion experiment at the LHC~\cite{ALICE:2008ngc,ALICE:2014sbx}.
The \pp dataset used in this analysis was collected in $2017$ at \five, while the \pPb dataset was collected in $2016$ at \fivenn, during the LHC Run $2$.
The analyzed data samples were collected with a minimum bias (MB) trigger and consist of $968\times10^{6}$ MB events for \pp collisions, corresponding to an integrated luminosity $L_{\rm pp}=18.9\pm 0.4$~nb$^{-1}$~\cite{ALICE:2018pqt}, and $624\times10^{6}$ MB events for \pPb collisions, corresponding to $L_{\rm pPb}=298\pm11$~$\mu$b$^{-1}$~\cite{ALICE:2014gvw}.
In \pPb collisions, a rapidity shift $\Delta y=0.465$ is needed in the direction of the proton beam to transform from the ALICE laboratory frame to the nucleon--nucleon center-of-mass frame due to the asymmetry of the colliding beam energies; protons at $4$~\TeV energy are collided into fully stripped $_{82}^{208}$Pb ions at $1.58$~\TeV per nucleon energy~\cite{ALICE:2014xsp}.

Events were triggered using the \VZERO detector~\cite{ALICE:2013axi}, which consists of two scintillator arrays located at forward and backward rapidity.
It covers the pseudorapidity regions $-3.7<\eta<-1.7$ (\VZEROC) and $2.8<\eta<5.1$ (\VZEROA).
To select the MB trigger, coincident signals are required in both the \VZEROA and \VZEROC detectors.
Beam-induced background events, such as beam--gas interactions or out-of-bunch pileup within the \VZERO detector readout time, are rejected offline by using the timing information from the \VZERO detectors and the number of reconstructed points and track segments in the Silicon Pixel Detector (\SPD), which are expected to be uncorrelated for background events.
The SPD equips the two innermost layers of the Inner Tracking System (ITS), a silicon tracker with six layers, and covers the pseudorapidity interval $\abs{\eta} < 1.4$ around midrapidity.
In-bunch pileup events, where multiple interactions occur in the same bunch crossing, are rejected by requiring that only a single primary vertex is reconstructed with the \SPD in the event~\cite{ALICE:2014sbx}.
For the data samples considered in this paper, pileup events amount to less than $1\%$ of the event sample both in \pp and \pPb collisions~\cite{ALICE:2021wct}.
Accepted events are required to have the reconstructed primary vertex position along the beam axis within $10$~cm from the center of the detector~\cite{ALICE:2014sbx}.

Charged-particle jets are reconstructed using tracks of primary-charged-particle candidates produced in the collision.
Primary charged particles are defined as all particles with a mean proper lifetime $\tau>1~{\rm cm}/c$ which are either produced directly in the interaction or from decays of particles with a mean proper lifetime $\tau<1~{\rm cm}/c$.
This excludes particles produced in interactions with the detector material and products of weak decays~\cite{ALICE:2017hcy}.
The charged-particle trajectories are reconstructed using information from the \ITS~\cite{ALICE:2010tia} and the Time Projection Chamber (\TPC)~\cite{Alme:2010ke}.
These detectors are located inside a large solenoidal magnet that provides a uniform magnetic field of $B=0.5$~T.
Tracks were selected with transverse momenta $\pTtrk>0.15$~\GeVc and in a pseudorapidity range $\abs{\eta}<0.9$ over the full azimuth $0 < \varphi < 2\pi$.

In order to achieve a uniform azimuthal angle distribution and the high-quality momentum resolution required for jet reconstruction, the charged track selection utilized a hybrid selection technique that compensates for local inefficiencies in the \SPD.
Two distinct classes of tracks are combined in the hybrid approach~\cite{ALICE:2013dpt}.
The first class consists of tracks that have at least one hit in the \SPD.
The second class contains tracks without hits in the \SPD, in which case the primary interaction vertex is used to constrain the trajectory in the track fit to improve the determination of their transverse momentum.
The charged track momentum resolution $\sigma(\pt)/\pt$ is estimated using the covariance matrix of the track fit~\cite{ALICE:2014sbx} and is approximately $0.8\%$ at $\pTtrk=1$~\GeVc and $4\%$ at $\pTtrk=50$~\GeVc.

Data corrections on instrumental effects were based on Monte Carlo (MC) simulations, which included a detailed description of the detector geometry and response, using the \Geant package~\cite{Brun:1082634}.
The simulations were performed using the \Pyeight event generator~\cite{Sjostrand:2014zea} with the Monash $2013$ tune~\cite{Skands:2014pea} for \pp collisions and the \Pysix~\cite{Sjostrand:2006za} with the Perugia $2011$ tune~\cite{Skands:2010ak} for \pPb collisions.
The simulated data were analyzed in the same way as the real data.


\section{Data analysis}%
\label{sec:dana}

\subsection{Jet reconstruction}

The strategy for the jet reconstruction closely followed the procedures used by the ALICE Collaboration in the Run $1$ analysis, including the background density estimation~\cite{ALICE:2015umm}.
Jet finding was performed using the FastJet $3.2.1$~\cite{Cacciari:2011ma} package.
Signal jets were reconstructed from charged-particle tracks using the anti-$\kT$ sequential clustering algorithm~\cite{Cacciari:2008gp} with resolution parameters $R = 0.2$, $0.3$, and $0.4$.
The four-momenta of the jet constituents were combined using the boost-invariant $\pt$ recombination scheme, treating the jet constituents as massless.
To ensure that jets were well contained in the \TPC acceptance, the pseudorapidity coverage of the reconstructed jets was constrained to $\abs{\etaj}<0.5$ for all jet resolution parameters.
The area of the jet was required to be $\Ajet>0.6\pi R^{2}$ to suppress the contribution from pure background jet clusters~\cite{ALICE:2019qyj}.
Jets which contained tracks with \pt larger than $100$~\GeVc were rejected, in order to suppress contamination by fake tracks and ensure good momentum resolution.
This selection rejects less than $1\%$ of jets in the considered kinematic region, and the impact on the raw jet spectrum is minor.

Reconstructed jets from the hard process are always accompanied by soft background that does not originate from the hard process, known as the underlying event.
The transverse momentum of selected signal jets in \pPb collisions was subsequently corrected for the average underlying event contribution~\cite{Cacciari:2007fd} according to the formula
\begin{equation}
\label{eq:pTcorr}
\pTjch=\pTjchraw-\rhoch\times\Ajet,
\end{equation}
where the transverse momentum density \rhoch of particles produced by the underlying event in \pPb collisions was estimated on an event-by-event basis using the so-called improved CMS method~\cite{CMS:2012rmf}
\begin{equation}
\label{eq:rho}
\rhoch = {\rm median}\left\{\frac{\pTj^{\kT}}{\Ajet^{\kT}}\right\}\times C.
\end{equation}
Here $\Ajet^{\kT}$ and $\pTj^{\kT}$ are the area and the transverse momentum of the jet clusters found using the \kT algorithm.
The $\kT$ jets had the same resolution parameter as the \akT jets, and the pseudorapidity range of the reconstructed $\kT$ jets spanned $\abs{\eta_{\rm jet}} < 0.9$.
The jet active area~\cite{Cacciari:2008gn} was estimated by distributing ghost particles into the $\eta$--$\varphi$ acceptance.
The ghost particle density is $200$ per unit area and corresponds to $0.005$ area per ghost particle.
The two highest-\pt jets in the event were excluded from the estimation of the background in order to suppress impact of physical jets on \rhoch~\cite{Cacciari:2010te}.
The scaling factor $C$ is used to account for regions without particles. It is defined as
\begin{equation}
\label{eq:corrFactor}
C=\frac{\sum_{\rm j} A_{\rm j}}{A_{\rm{acc}}}.
\end{equation}
Here $A_{\rm j}$ is the area of each \kT jet with at least one real track (i.e. excluding ghosts), and $A_{\rm{acc}}$ is the area of charged-particle acceptance.
In \pp collisions, the underlying event constitutes approximately $\pt=1$~\GeVc per jet, and was not subtracted from the raw jet spectrum~\cite{ALICE:2022hyz}.

\subsection{Corrections}
\label{sec:corr}

The measured jet spectrum is distorted predominantly due to the finite detector resolution and local background fluctuations with respect to the mean underlying event density.
A correction procedure known as unfolding was used to correct for these effects~\cite{Blobel:2011fih, Cowan:2002in}.
The response matrices for these two effects are determined separately and combined by making a product of the two~\cite{ALICE:2015mjv}. 

The response matrix describing jet momentum smearing due to instrumental effects was determined from the \Pythia MC simulation.
Detector-level jets were reconstructed by transporting the generated particles through a full simulation of the ALICE detector using the \Geant transport model~\cite{Brun:1119728}.
These jets were geometrically matched to the corresponding particle-level jets on a jet-by-jet basis by requiring that the angular distance $\Delta R=\sqrt{(\Delta\eta)^2+(\Delta\varphi)^2}<0.6R$, where $\Delta\eta$ and $\Delta\varphi$ are the differences in pseudorapidity and azimuthal angle between the detector-level and particle-level jets.
The particle-level jets were reconstructed without imposing any selection on the constituents.
Consequently, during the unfolding process, the reconstructed jets are corrected to a constituent-charged-particle momentum of $0$~\GeVc.

The probability of a particle-level jet not matching any detector-level jet is considered as the jet reconstruction efficiency, is applied to correct the unfolded jet spectrum.
Its value is about $80\%$ at jet $\pt\approx 10$~\GeVc and increases to $99\%$ for jets with $\pt > 30$~\GeVc in the considered kinematic region.
Additionally, the probability that a detector-level jet is falsely matched to a non-corresponding particle-level jet is called the jet reconstruction purity.
The jet reconstruction purity is found to be above $99\%$ in the measured kinematic range.
Hence, the raw jet spectrum was not corrected for the reconstruction purity.

The performance of the jet reconstruction was assessed using the MC simulation.
Two variables are evaluated: the shift of the mean jet energy scale (JES) $\Delta_{\rm JES}=\avg{(\pTjdet-\pTjtruth)/\pTjtruth}$ and the jet energy resolution ${\rm JER}=\sigma(\pTjdet)/\pTjtruth$, where $\pTjdet$ and $\pTjtruth$ are the transverse momenta of the measured jet and the corresponding truth jet, and $\sigma(\pTjdet)$ denotes the width of the $\pTjdet-\pTjtruth$ distribution as a function of $\pTjtruth$.
The $\Delta_{\rm JES}$ distribution is asymmetric and has a long negative tail due to the reconstruction inefficiency and a sharp peak centered around $\pTjtruth=\pTjdet$.
The most probable scenario is that the measured $\pTjch$ is close to the jet \pt at particle level.
The $\Delta_{\rm JES}$ distribution has a mean value of $-15\%~(-20\%)$ and $-27\%~(-30\%)$ for the \pTjtruth interval $20<\pTjtruth<30$~\GeVc and $100<\pTjtruth<120$~\GeVc, respectively, for \pp~(\pPb) collisions.
The value of the JER varies from $23\%~(20\%)$ to $27\%~(38\%)$ in \pp~(\pPb) collisions at $\pTjtruth=20$~\GeVc and $\pTjtruth=150$~\GeVc, respectively.
Both $\Delta_{\rm JES}$ and JER exhibit a weak $R$ dependence.

The response matrix which accounts for the smearing due to local background fluctuations, was obtained with the random cone (RC) method~\cite{ALICE:2012nbx}.
Cones with a resolution parameter $R_{\rm cone}$ equal to that of the jet were placed randomly in the $\eta-\varphi$ space in each event, within the \ITS and \TPC acceptance.
The background fluctuations were evaluated by comparing the sum of the $\pt$ of tracks inside the cone, $\pt^{\rm RC}$, with the expected average contribution due to the underlying event as follows:
\begin{equation}
\label{eq:deltapT}
\delta\pt^{\rm RC}=\pt^{\rm RC}-\rhoch\pi R^{2}_{\rm cone}.
\end{equation}
In this study, two definitions for the random cone were considered.
First, the random cones were required not to overlap with the leading and subleading jets in an event.
Second, the cones were placed in a perpendicular direction to the leading jet in an event.
These two approaches yield a consistent result and their difference was considered as a source of systematic uncertainty.
The nominal result used the $\delta\pt^{\rm RC}$ matrix obtained by the first approach.
The corresponding $\delta\pt^{\rm RC}$ distribution has a width of $\sigRC = 2.01$~\GeVc for $R = 0.2$, $3.01$~\GeVc for $R = 0.3$, and $4.01$~\GeVc for $R = 0.4$.
The response matrix for the local background fluctuations was constructed row-by-row by taking the $\delta\pt^{\rm RC}$ distribution and shifting it along the \pTjdet axis by the amount \pTjtruth corresponding to each row~\cite{ALICE:2015mjv}.
In \pp collisions, no correction for the background fluctuations was applied, and the raw spectrum was corrected for the instrumental effects only.

In our case the use of an unregularized unfolding method led to a non-physically behaving solution, which wildly oscillated due to statistical fluctuations of the measured spectrum.
To suppress these fluctuations, unfolding algorithms which impose a smoothness criterion were used in this analysis.
The regularization procedure introduces additional systematic uncertainties on the unfolded spectrum, as discussed in Sec.~\ref{sec:shapeunc}.
The singular value decomposition (SVD-) based method~\cite{Hocker:1995kb} was used to perform the unfolding correction for the primary analysis.
The optimum regularization parameter, $k_{\rm reg}$, was chosen such that it corresponds to the point at which statistically significant components of the $d$-vector distribution change to noise.
The values of $k_{\rm reg}=4$ and $6$ are adopted for \pp and \pPb collisions, respectively.
The unfolding was also performed using the iterative Bayesian method~\cite{DAgostini:1994fjx}.
For the regularization of the Bayesian unfolding, convergence was determined by the stability of the unfolded solution to successive iterations, while minimizing the statistical uncertainties~\cite{ALICE:2017svf}.
The convergence was achieved after the second iteration for both \pp and \pPb collisions.
These unfolding algorithms are implemented in the RooUnfold framework~\cite{Adye:2011gm}.
The difference between the two unfolding methods was assigned as a source of systematic uncertainty.
Input to the unfolding was the raw jet spectrum measured in the range $10<\pTjch<140$~\GeVc for \pp collisions and $10<\pTjch<160$~\GeVc for \pPb collisions.
The fraction of particle-level jets that are reconstructed outside the measured range, referred as the kinematic efficiency, is applied to the unfolded spectrum in addition to the jet reconstruction efficiency correction.
The kinematic efficiency is above $80\%$ in the measured range.
The unfolding correction required a prior spectrum as the starting point of the algorithm.
In this analysis, by default the prior spectrum was defined by the particle-level distribution generated using the \Pythia simulation.
The unfolding correction of the raw jet spectrum ranges from $30\%$ to $60\%$ across the analyzed \pt range.

The robustness of the unfolding procedure and the mathematical validity of the unfolded solution were established through refolding and closure tests as done in the previous ALICE jet measurements~\cite{ALICE:2017svf}.
In the refolding test, the unfolded spectrum was multiplied by the response matrix and the resulting distribution was compared to the raw jet spectrum.
The closure test involved the use of two statistically independent MC samples.
The first sample was utilized to construct the response matrix and provide the prior distribution.
The second one served to generate pseudo-data, which were obtained by resampling the central values using Poissonian smearing. 
The pseudo-data was then unfolded in the same way as the real data and compared to the particle-level distribution.
Within statistical uncertainties, the solutions from both tests were able to recover the input distribution.

Statistical uncertainties of the unfolded solutions were evaluated based on pseudo-random experiments.
In this approach, the bin contents of the input measured spectrum were smeared according to given statistical uncertainties obtaining an ensemble of randomized spectra.
The unfolding was then applied to each of these spectra and the resulting statistical uncertainty in each bin was obtained from a covariance matrix corresponding to the ensemble.
To cross-check the correlation of the statistical uncertainties of the bin contents of the jet spectrum, the statistical uncertainties were also evaluated using the Bootstrap method~\cite{Efron:1979bxm, ATLAS:2017ble} and found to be consistent with the pseudo-random experiments.
When calculating the ratio of jet cross sections, the spectra which appear in the numerator and denominator are from the same input data.
The degree of correlation between the same $\pt$ bins is nevertheless suppressed since jets reconstructed with different $R$ may fall in different $\pt$ bins.
Thus the correlation was not taken into account; the numerator and denominator are treated as independent, which leads to a conservative estimate of the statistical uncertainties.

\subsection{Systematic uncertainties}

The systematic uncertainties of the \pt-differential charged-particle jet cross section and $\RpPbjch$ are quantified by varying several parameters with respect to the primary analysis.
The uncertainties are categorized based on their point-to-point correlation into correlated uncertainty, shape uncertainty, and normalization uncertainty.
The correlated uncertainty is positively correlated among all the \pTjch bins.
It includes the uncertainty on the tracking efficiency and the uncertainty on the jet momentum smearing due to local background fluctuations.
The shape uncertainty is the uncertainty which is anti-correlated between parts of the unfolded spectrum, which affects the shape of the final \pTjch spectrum.
It arises mainly due to assumptions in the unfolding procedure.
The normalization uncertainties on the luminosity measurement, as described in Sec.~\ref{sec:setup}, were determined to be $2.34\%$~\cite{ALICE:2018pqt} and $3.7\%$~\cite{ALICE:2014gvw} for \pp and \pPb collisions, respectively.

The influence of the statistical fluctuations on the systematic uncertainties of the raw spectrum was suppressed by using pseudo-experiments as done in Refs.~\cite{ALICE:2021wct, ALICE:2017svf}.
For each source of uncertainty, several randomized instances of the raw jet \pTjch spectrum were generated by variations around the measured central value in each bin using a Gaussian distribution, with $\sigma$ taken to be the uncorrelated statistical error in the bin.
Each randomized instance was analyzed using (i) corrections for the primary analysis, and (ii) corrections that include the systematic variation.
For each randomized instance, the ratio of corrected jet \pTjch spectra resulting from (ii) and (i) was formed.
The systematic uncertainty in each \pTjch bin was defined as the mean value of the distribution of ratios obtained from all randomized instances.
The uncertainties were taken as symmetric and the total uncertainty for each category was obtained by making a quadratic sum of the uncertainties corresponding to individual sources.
The summary of the relative systematic uncertainties discussed in this section is presented in Table~\ref{tab:systunc}.

\subsubsection{Correlated uncertainties}

The main sources of correlated uncertainties are described below.

\begin{itemize}
\item[$\bullet$] Tracking efficiency: The dominant systematic uncertainty arises from the uncertainty on the ALICE tracking efficiency, which composes of two parts.
The first part was estimated by simultaneously varying track selection criteria in the \TPC in data and the MC simulation.
The second one was determined by the discrepancy in the \TPC-\ITS track matching efficiency between data and simulations.
The uncertainty on the inclusive \pt spectrum of charged particles was found to be $3\%$ in \pp collisions~\cite{ALICE:2019wqv}.
In \pPb collisions, the value was found to increase with $\pt$ from $1\%$ at low \pt ($\sim 0.5$~\GeVc) up to $2.5\%$ at high \pt ($\sim 14$~\GeVc).
To assess how this uncertainty impacts the charged-particle jet spectrum, the unfolding was also performed with a response matrix which accounted for the lower track reconstruction efficiency.
The difference between the jet spectrum obtained with the modified response matrix and the default one is adopted as the uncertainty on tracking efficiency.

\item[$\bullet$] \Pythia fragmentation: The instrumental response matrix used in the unfolding correction was built using \Pythia simulations. 
The impact of the hadronization model in \Pythia was assessed using the jet angularity observable.
The jet angularity is defined as $g = \sum_{i} (p_{{\rm T},i} \times r_{i}) / \pTjch$, where $p_{{\rm T},i}$ is the \pt of the $i^{\rm th}$ constituent of the charged-particle jet and $r_{i} = \sqrt{\Delta\eta_{i}^{2} + \Delta\varphi_{i}^{2}}$ is the distance of the $i^{\rm th}$ constituent from the jet axis at the particle level.
The angularity distribution was simulated using \Pyeight and \Hrwg~\cite{Bahr:2008pv,Bellm:2015jjp} simulations.
The corresponding jet angularity distributions differ up to about $30\%$.
To assess the uncertainty associated with the detector effects on the fragmentation model, the instrumental response matrix was re-weighted according to the jet angularity.
Specifically, the instrumental response matrix was re-weighted such that the $50\%$ largest angularity jets were weighted an additional $\pm 30\%$ relative to the $50\%$ lowest angularity jets~\cite{ALICE:2019qyj}.
The modified response matrix was then used to unfold the measured spectrum.
The difference between the unfolded jet spectrum obtained with the re-weighted response matrix and the default one is taken as the uncertainty.

\item[$\bullet$] Background fluctuations: In \pPb collisions, the $\delta p_{\rm T}$ matrix was constructed with cones perpendicular to the leading jet, as described in Section~\ref{sec:corr}.
The uncertainty on background fluctuations is estimated by taking the difference between the resulting jet spectrum and the default one.
\end{itemize}

\subsubsection{Shape uncertainties}
\label{sec:shapeunc}

The main sources of the shape uncertainties are described below.

\begin{itemize}
\item[$\bullet$] Variation of the unfolding algorithm: The \pTjch spectrum was unfolded with the iterative Bayesian unfolding method.
\item[$\bullet$] Variation of the regularization parameter: The regularization parameter $k_{\rm reg}$ in the SVD-based unfolding was varied by $\pm1$ with respect to the optimal value.
\item[$\bullet$] Variation of the prior: The prior spectrum was changed to a \pt spectrum of jets calculated with \Pwgpy simulations.
\item[$\bullet$] Variation of the lower \pt spectrum cutoff: The minimum \pt of the measured jet spectrum used in the unfolding correction was required to be greater than $\sigRC$ of the $\delta\pt^{\rm RC}$ distribution.
The sensitivity of the unfolded result to combinatorial jets was tested by varying the lower range of the measured jet spectrum by $\pm3$~\GeVc.
\end{itemize}

The systematic uncertainties on the cross section ratios between jet spectra obtained with different $R$ values are determined using the same strategy as in previous ALICE measurements~\cite{ALICE:2019qyj}.
The numerator and denominator were varied simultaneously and compared to the default jet cross section ratio.

The shape uncertainties between the \pp and \pPb collision systems are considered uncorrelated and are fully propagated to the $\RpPbjch$.
Due to the partial correlation between the tracking efficiency uncertainties in \pp and \pPb collisions~\cite{ALICE:2015umm}, the tracking efficiency uncertainty on the \RpPbjch is considered to be the maximum uncanceled part between the two collision systems.

\begin{sidewaystable}
      \centering
      \caption{Summary of the contributions to the relative systematic uncertainty for the charged-particle jet cross section in \pp and \pPb collisions, and $\RpPbjch$ for $R=0.2$, $0.3$, and $0.4$.
      The uncertainties depend on the \pTjch and the table shows representative values corresponding to the first and last \pTjch interval for each jet resolution parameter.
      The contributions are assumed to be independent and are summed in quadrature, resulting in the total uncertainty. See text for details.}
        \begin{tabular}{cccccccccccccccc}
        \toprule
        \toprule
              & \multirow{2}[4]{*}{$R$} & \multirow{2}[4]{*}{\pTjch (\GeVc)} &       & \multicolumn{4}{c}{Correlated uncertainty (\%)} &       & \multicolumn{5}{c}{Shape uncertainty  (\%)} &       & \multirow{2}[4]{*}{Norm.  (\%)} \\
    \cmidrule{5-8}\cmidrule{10-14}          &       &       &       & Trk. eff. & Frag. & Bkg. fluc. & Total  &       & Algo. & Reg. par. & Prior & Bin trunc. & Total &       &  \\
        \midrule
        \midrule
        \multirow{6}[6]{*}{\pp} & \multirow{2}[2]{*}{$0.2$} & $10$~--~$20$ &       & $5.3$  & $2.1$  & \textemdash & $5.7$  &       & $0.06$  & $0.07$  & $0.15$  & $0.47$  & $0.50$  &       & \multirow{6}[6]{*}{$2.34$} \\
              &       & $120$~--~$140$ &       & $9.7$  & $1.1$  & \textemdash & $9.8$  &       & $8.0$  & $4.3$  & $0.70$  & $4.3$  & $10$ &       &  \\
    \cmidrule{2-3}\cmidrule{5-8}\cmidrule{10-14}          & \multirow{2}[2]{*}{$0.3$} & $10$~--~$20$ &       & $6.3$  & $1.5$  & \textemdash & $6.5$  &       & $0.04$  & $0.05$  & $0.43$  & $0.51$  & $0.67$  &       &  \\
              &       & $120$~--~$140$ &       & $9.6$  & $0.40$  & \textemdash & $9.6$  &       & $9.7$  & $4.1$  & $1.1$  & $2.7$  & $11$ &       &  \\
    \cmidrule{2-3}\cmidrule{5-8}\cmidrule{10-14}          & \multirow{2}[2]{*}{$0.4$} & $10$~--~$20$ &       & $7.3$  & $2.4$  & \textemdash & $7.7$  &       & $0.08$  & $0.12$  & $0.06$  & $0.44$  & $0.47$  &       &  \\
              &       & $120$~--~$140$ &       & $10$ & $0.40$  & \textemdash & $10$ &       & $9.3$  & $7$  & $0.64$  & $1.2$  & $12$ &       &  \\
        \midrule
        \multirow{6}[6]{*}{\pPb} & \multirow{2}[2]{*}{$0.2$} & $10$~--~$20$ &       & $16$ & $0.70$  & $0.21$  & $16$ &       & $2.0$  & $0.34$  & $0.22$  & $0.21$  & $2.05$  &       & \multirow{6}[6]{*}{$3.70$} \\
              &       & $140$~--~$160$ &       & $4.6$  & $0.80$  & $1.3$  & $4.8$  &       & $11$ & $3.9$  & $1.4$  & $3.9$  & $12$ &       &  \\
    \cmidrule{2-3}\cmidrule{5-8}\cmidrule{10-14}          & \multirow{2}[2]{*}{$0.3$} & $10$~--~$20$ &       & $16$ & $1.4$  & $0.54$  & $16$ &       & $0.51$  & $0.78$  & $0.24$  & $0.54$  & $1.1$  &       &  \\
              &       & $140$~--~$160$ &       & $7.7$  & $1.5$  & $0.36$  & $7.9$  &       & $8.8$  & $4.2$  & $1.7$  & $1.9$  & $10$ &       &  \\
    \cmidrule{2-3}\cmidrule{5-8}\cmidrule{10-14}          & \multirow{2}[2]{*}{$0.4$} & $10$~--~$20$ &       & $16$ & $1.7$  & $1.2$  & $16$ &       & $3.9$  & $0.74$  & $0.19$  & $0.20$  & $4.0$  &       &  \\
              &       & $140$~--~$160$ &       & $9.3$  & $0.60$  & $0.16$  & $9.3$  &       & $3.3$  & $3.3$  & $3.2$  & $3.3$  & $6.6$  &       &  \\
        \midrule
        \multirow{6}[6]{*}{\RpPbjch} & \multirow{2}[2]{*}{$0.2$} & $10$~--~$20$ &       & $5.3$  & $2.2$  & $0.21$  & $5.8$  &       & $2.0$  & $0.35$  & $0.55$  & $0.51$  & $2.2$  &       & \multirow{6}[6]{*}{$4.37$} \\
              &       & $120$~--~$140$ &       & $4.8$  & $1.4$  & $1.3$  & $5.2$  &       & $11$ & $4.9$  & $3.1$  & $4.9$  & $13$ &       &  \\
    \cmidrule{2-3}\cmidrule{5-8}\cmidrule{10-14}          & \multirow{2}[2]{*}{$0.3$} & $10$~--~$20$ &       & $6.3$  & $2.1$  & $0.54$  & $6.7$  &       & $0.51$  & $0.78$  & $0.49$  & $0.75$  & $1.3$  &       &  \\
              &       & $120$~--~$140$ &       & $7.3$  & $1.6$  & $0.36$  & $7.5$  &       & $12$ & $5.7$  & $1.1$  & $2.8$  & $14$ &       &  \\
    \cmidrule{2-3}\cmidrule{5-8}\cmidrule{10-14}          & \multirow{2}[2]{*}{$0.4$} & $10$~--~$20$ &       & $7.3$  & $3.0$  & $1.2$  & $8.0$  &       & $3.9$  & $0.75$  & $0.20$  & $0.49$  & $4.0$  &       &  \\
              &       & $120$~--~$140$ &       & $9.1$  & $0.72$  & $0.16$  & $9.1$  &       & $10$ & $7.2$  & $1.6$  & $2.2$  & $13$ &       &  \\
        \bottomrule
        \bottomrule
        \end{tabular}%
      \label{tab:systunc}%
    \end{sidewaystable}%


\section{Results}%
\label{sec:res}
\subsection{Inclusive charged-particle jet production cross section in \pp and \pPb collisions}

The charged-particle jet cross sections are reported differentially in \pTjch and \etaj as
\begin{equation}\label{eq:xsec}
\frac{\dd^{2}\sigma}{\dd\pTjch\dd\etaj}=\frac{1}{L}\frac{\dd^{2}{N}}{\dd\pTjch\dd\etaj},
\end{equation}
where $\dd^{2}{N} / \dd\pTjch\dd\etaj$ is the fully corrected $\pt$- and $\eta$-differential charged-particle jet yield.
The integrated luminosity for minimum bias events is denoted by $L$, see Sec.~\ref{sec:setup}.

\begin{figure}[!t]
\centering
\includegraphics[width=.96\textwidth]{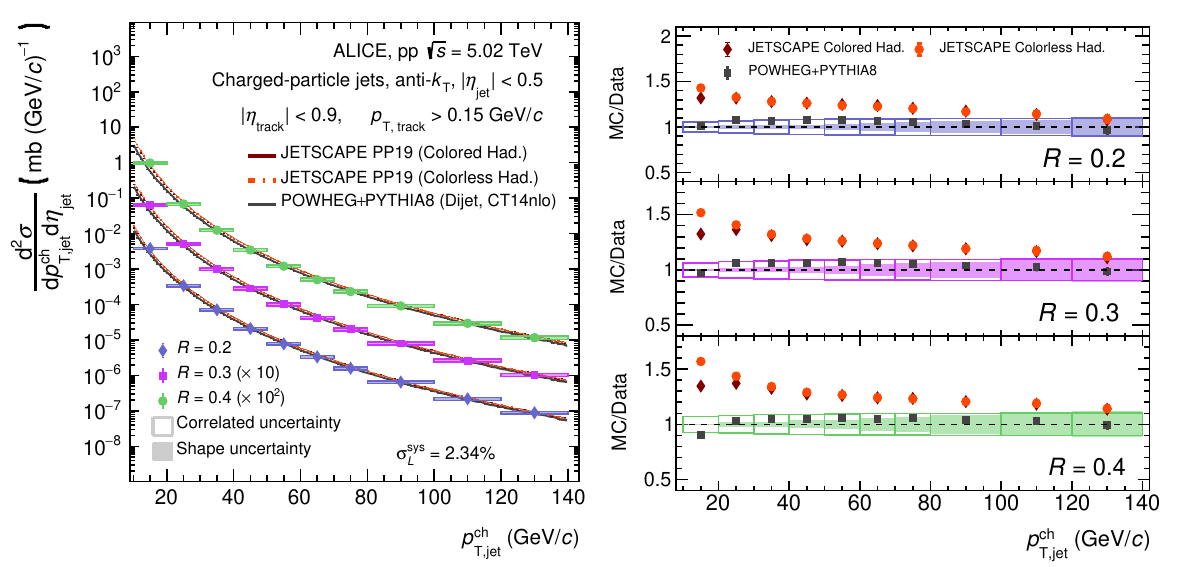}
\caption{Left panel: Fully corrected charged-particle jet cross section in \pp collisions at \five for $R=0.2$, $0.3$, and $0.4$ (scale factors as indicated are used for better visibility).
Statistical uncertainties are shown as vertical bars and are typically smaller than the marker size for all points.
Boxes indicate systematic uncertainties.
The additional normalization uncertainty due to luminosity is quoted separately as $\sigma_{L}^{\rm sys}$.
The data are compared to the predictions by \Pwgpy and \Scape.
Right panels: Ratio of the theory predictions to the measured data.
The correlated and shape systematic uncertainties on the ratio account for the data uncertainties and do not show the normalization uncertainty.}
\label{fig:xsecpp}
\end{figure}

\begin{figure}[!t]
\centering
\includegraphics[width=.96\textwidth]{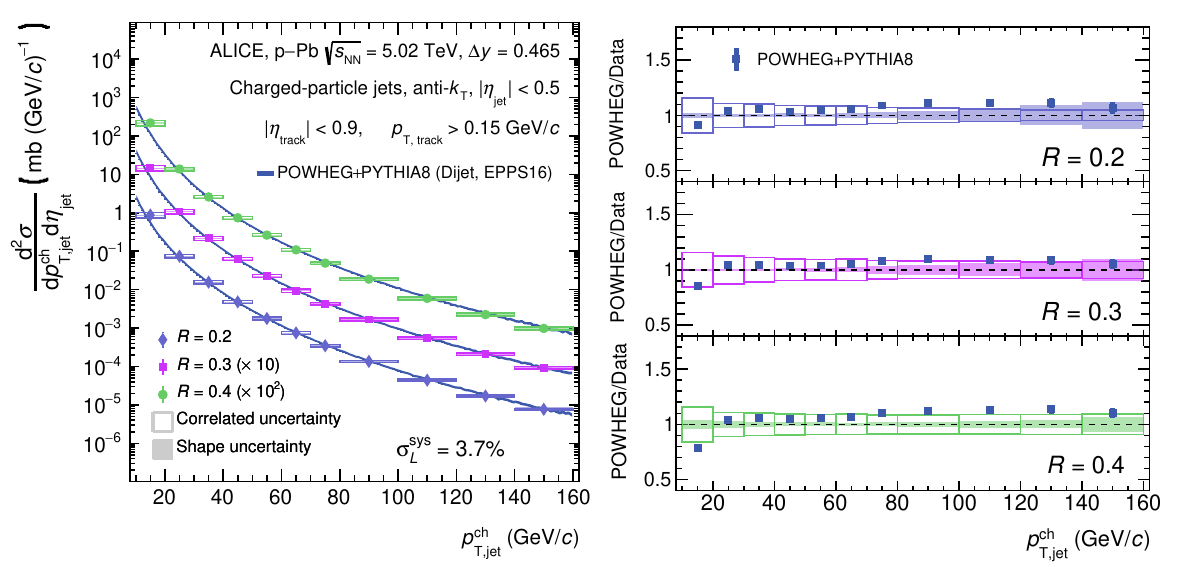}
\caption{Similar results as Fig.~\ref{fig:xsecpp} for \pPb collisions at \fivenn.}
\label{fig:xsecpPb}
\end{figure}

The fully corrected charged-particle jet cross sections for $R=0.2$, $0.3$, and $0.4$ in \pp and \pPb collisions are shown in Figs.~\ref{fig:xsecpp} and~\ref{fig:xsecpPb}, respectively.
The jet cross sections for larger jet resolution parameters are scaled by arbitrary factors described in the legend for better visibility.
These results are compatible with the previous results from ALICE~\cite{ALICE:2019wqv, ALICE:2015umm} and cover a wider \pTjch interval, and are consistent with the ALICE charged-particle jet measurements~\cite{ALICE:2023waz}.
The measurements are compared to two theoretical predictions: 

\begin{itemize}
\item[$\bullet$] \Pwgpy: The matrix elements are computed at NLO accuracy with the \Pwg method~\cite{Frixione:2007vw, Nason:2004rx} using the dijet process~\cite{Alioli:2010xa} implemented in the \Pwgbox V$2$ framework~\cite{Alioli:2010xd}.
The \Pwgbox is interfaced with the \Lha package~\cite{Buckley:2014ana} to supply different (n)PDF sets for describing the non-perturbative initial states, and \Pyeight~\cite{Sjostrand:2014zea} tune A$14$~\cite{ATL-PHYS-PUB-2014-021} for parton shower and fragmentation.
The simulation employs the CT$14$ NLO~\cite{Dulat:2015mca}, TUJU$21$ NNLO~\cite{Helenius:2021tof}, CTEQ$6.1$ NLO~\cite{Stump:2003yu}, and nNNPDF3.0 NLO~\cite{AbdulKhalek:2022fyi} free-proton PDF sets.
In the case of \pPb collisions, the corresponding nPDF sets EPPS$16$ NLO~\cite{Eskola:2016oht}, nTUJU$21$ NNLO~\cite{Helenius:2021tof}, nCTEQ$15$WZSIH NLO~\cite{Kovarik:2015cma,Kusina:2020lyz}, and nNNPDF$3.0$ NLO~\cite{AbdulKhalek:2022fyi} are used, accounting also for the shift in rapidity.
The default values of $\alpha_{\rm s}$, renormalization, and factorization scales are adopted in the results shown in this paper.

\item[$\bullet$] \Scape: \Pyeight~\cite{Sjostrand:2014zea} is used to generate the initial hard scattering and the underlying event.
The intermediate shower is handled by the MATTER~\cite{Majumder:2013re, Cao:2017qpx} model that includes parton virtuality. After parton shower, QCD strings are formed through either a colored or a colorless hadronization scheme. The strings are subsequently fed into \Pyeight for string fragmentation~\cite{Andersson:1983ia,Andersson:1997xwk}.
The \Scape configuration used in this paper is referred to as the PP$19$ tune~\cite{JETSCAPE:2019udz} as implemented in \Scape V$3.4.1$~\cite{Putschke:2019yrg}.
The predictions are only shown for \pp collisions.
\end{itemize}

As shown in the right panels of Figs.~\ref{fig:xsecpp} and~\ref{fig:xsecpPb}, within the uncertainties of the data, the \Pwgpy predictions describe the data well, except for the lowest $\pTjch$ interval $10$--$20$~\GeVc where a maximum discrepancy of $\sim20\%$ is observed in both \pp and \pPb collisions.
The \Scape prediction in \pp collisions overestimates the data by $\sim50\%$ at low $\pt$.
The magnitude of the discrepancy depends on the jet $\pt$ and the resolution parameter $R$.
The low-$\pt$ region is sensitive to non-perturbative effects such as the initial state radiation and multiparton interactions.
It is not clear to which extent can these discrepancies be explained by improper modelling of these effects.

\begin{figure}[!t]
\centering
\includegraphics[width=.48\textwidth]{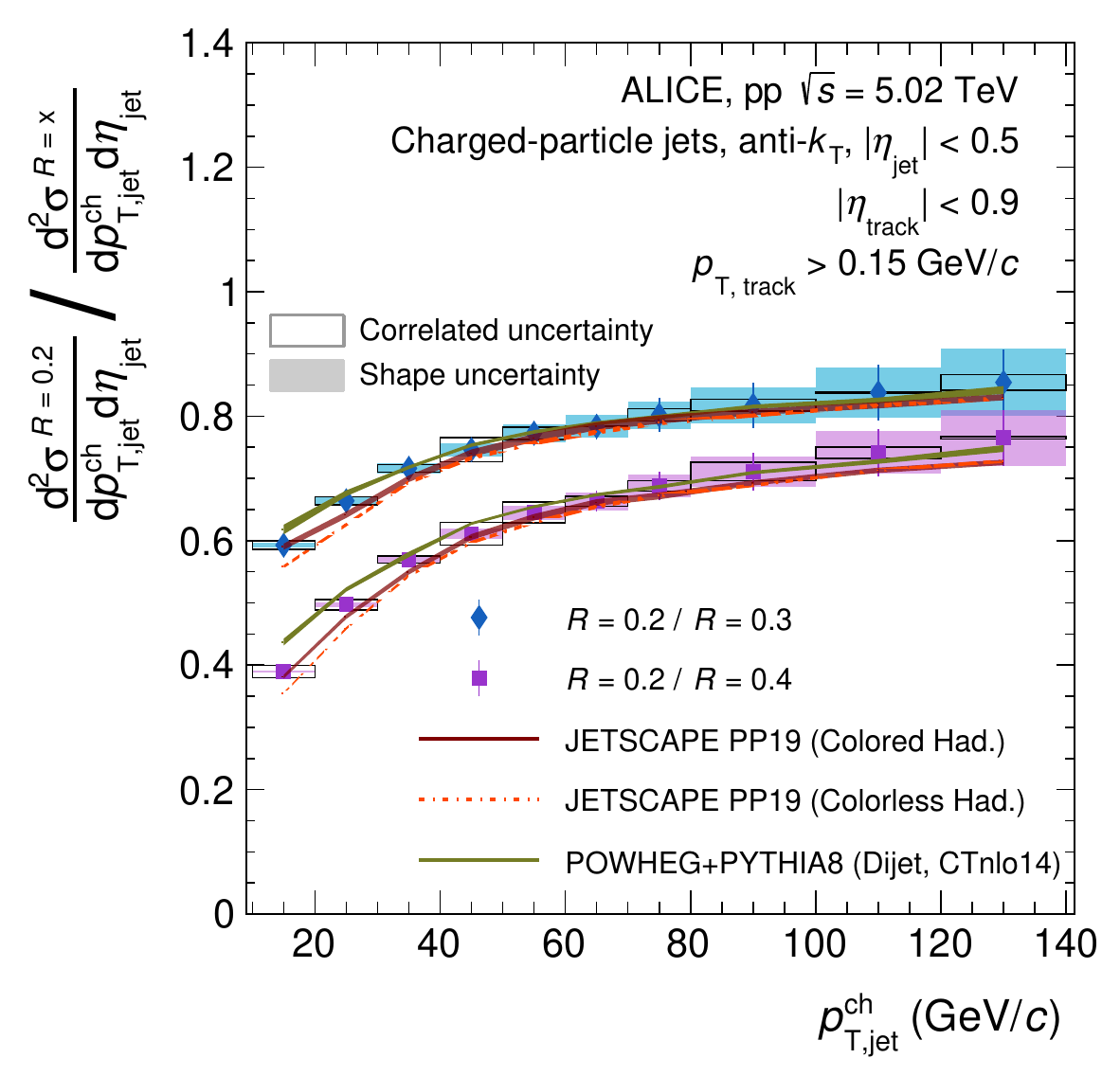}
\includegraphics[width=.48\textwidth]{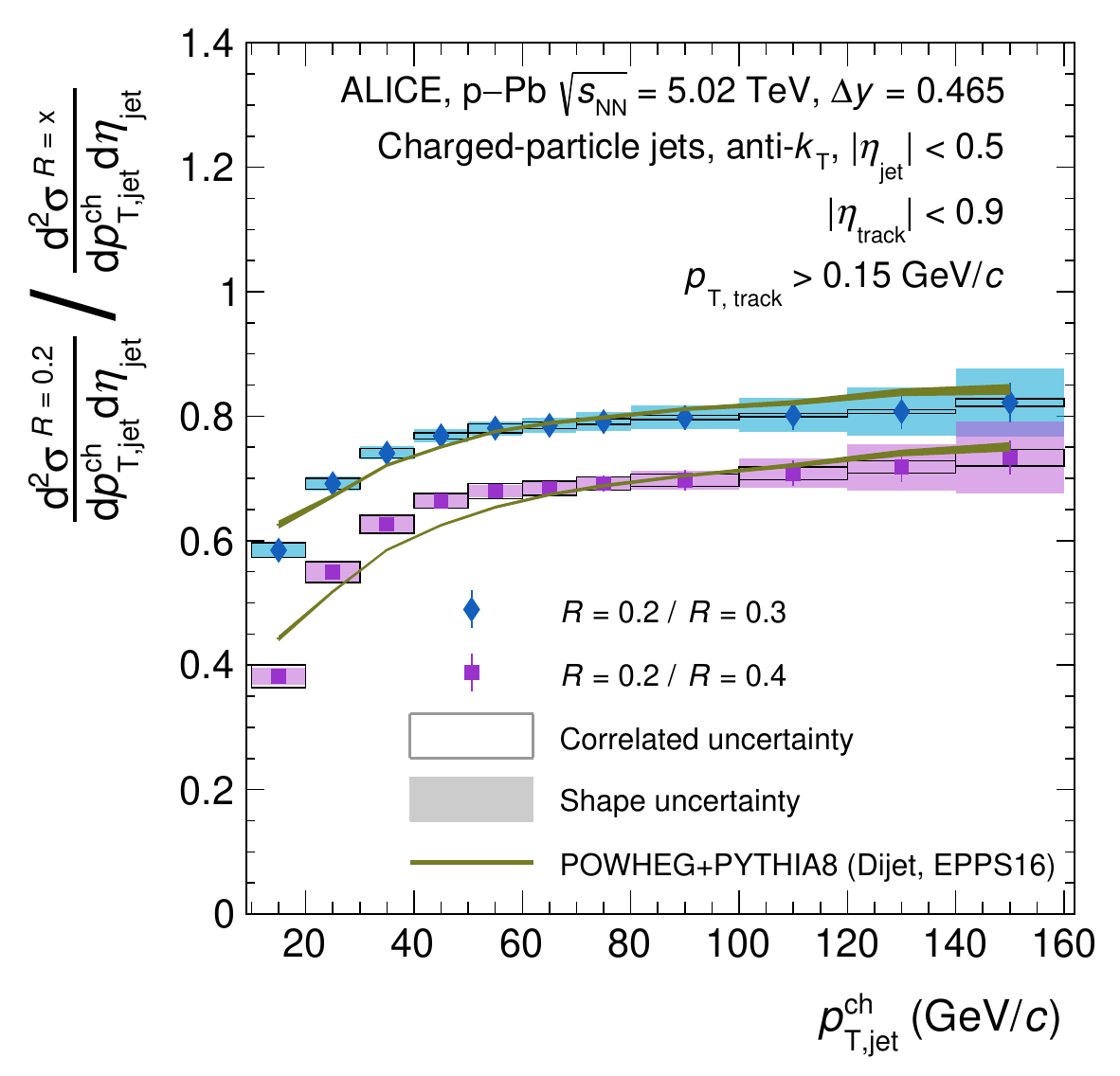}
\caption{Fully corrected charged-particle jet cross section ratios with $R=0.2$ to other jet resolution parameters.
The plots also include comparison to the \Pwgpy and \Scape (in case of \pp collisions) predictions.
Left panel: \pp collisions at \five.
Right panel: \pPb collisions at \fivenn.}
\label{fig:xsecratio}
\end{figure}

\begin{figure}[!t]
\centering
\includegraphics[width=.48\textwidth]{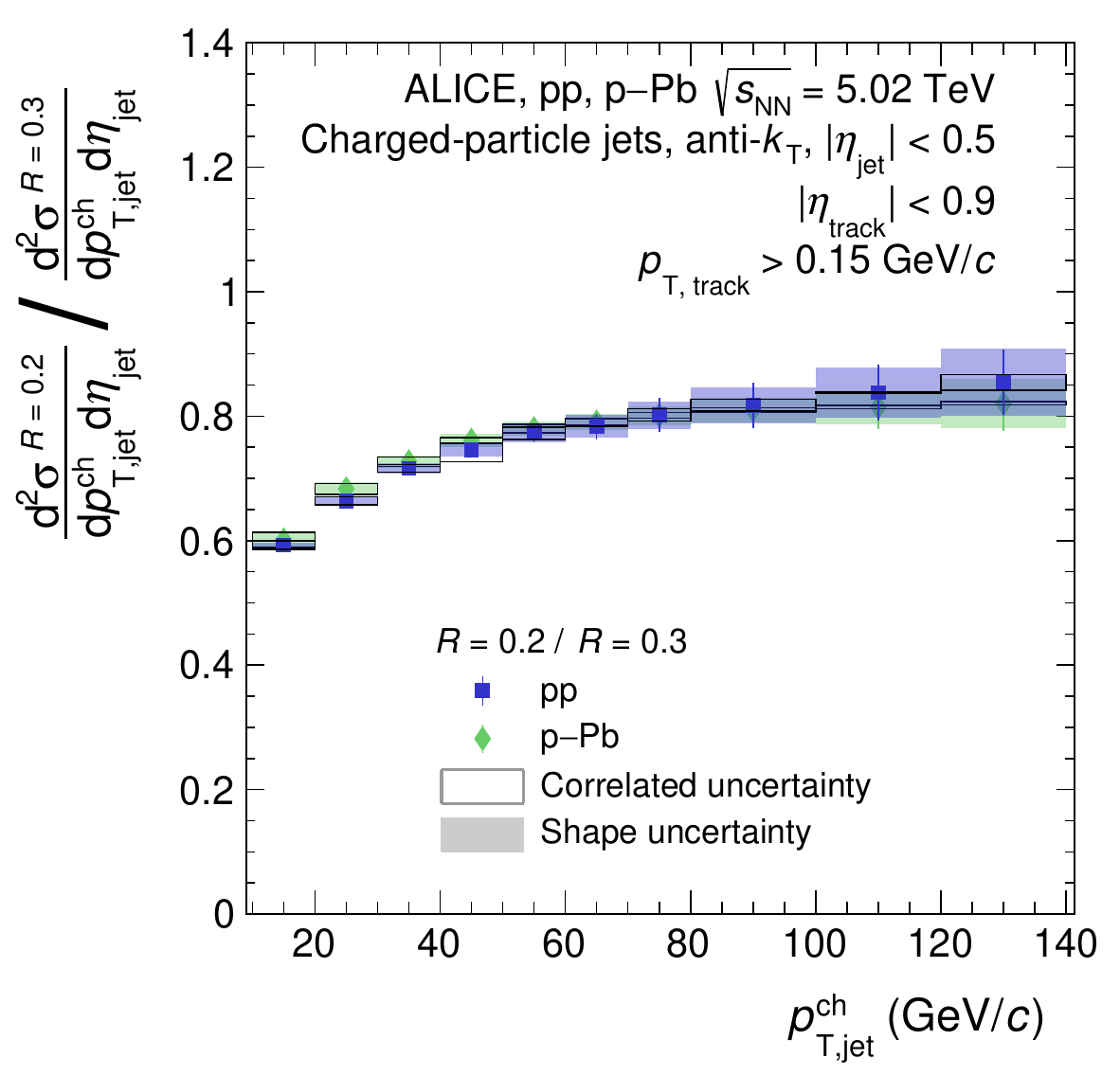}
\includegraphics[width=.48\textwidth]{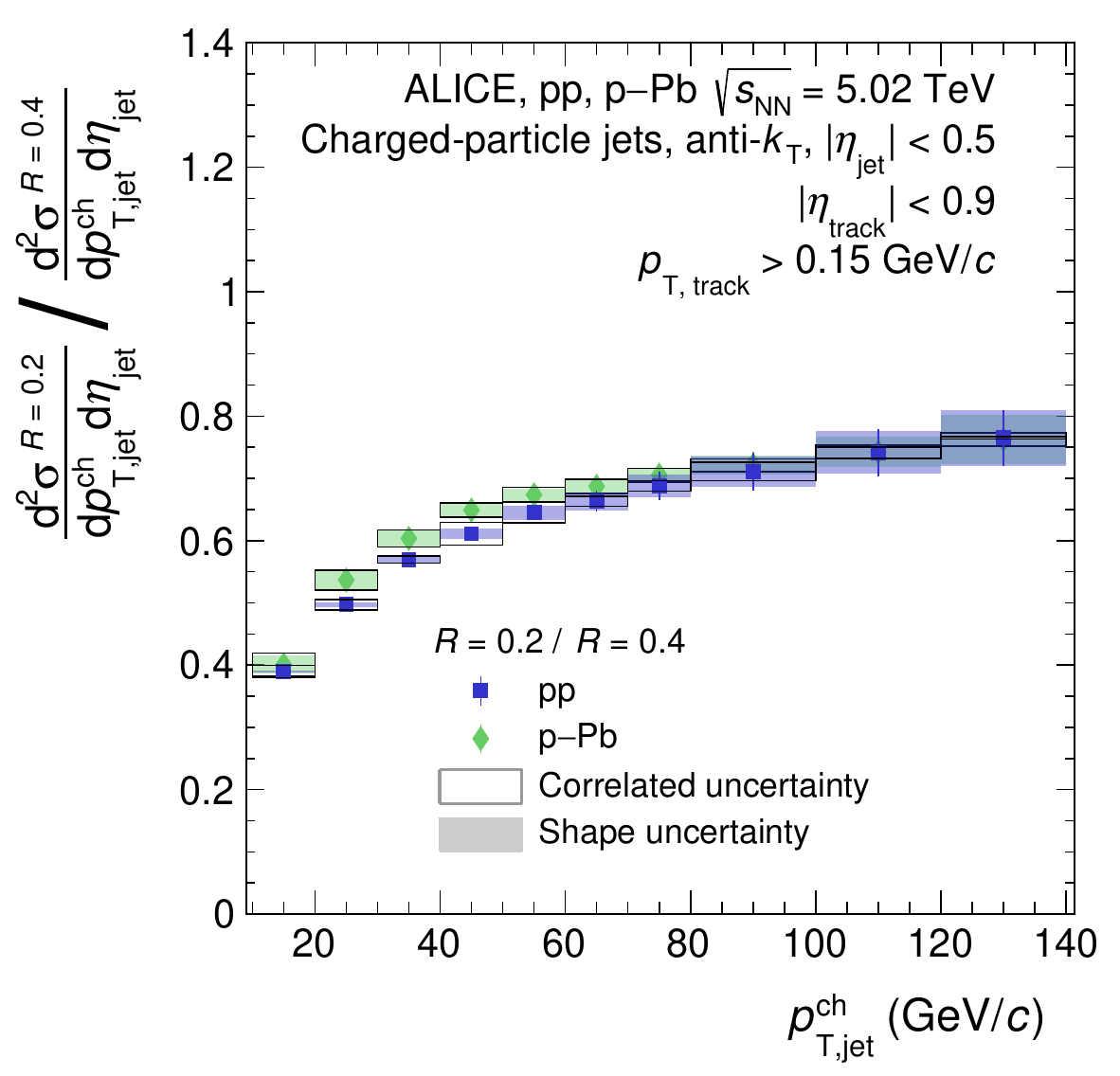}
\caption{Comparison of jet cross section ratios between \pp collisions and \pPb collisions at \fivenn.
The left panel shows the ratios between jets with $R=0.2$ and $R=0.3$, the right panel shows the ratios between jets with $R=0.2$ and $R=0.4$.}
\label{fig:xsecratioppVspPb}
\end{figure}

The jet cross section ratios were evaluated by dividing the spectrum with $R = 0.2$ by those with other resolution parameters as shown in Fig.~\ref{fig:xsecratio}.
The left panel is for \pp collisions, and the right panel is for \pPb collisions.
In both \pp and \pPb collisions, the data are compared with the \Pwgpy predictions.
General agreement between data and \Pwgpy predictions is observed within uncertainties.
In \pp collisions, the comparison between data and \Scape predictions is also presented.
\Scape predictions are consistent with data within uncertainties as well, but they show a difference from the \Pwgpy predictions at low $\pTjch$ in $20$--$40$~\GeVc.
The cross section ratio is sensitive to the collimation of particles around the jet axis and serves as an indirect measure of the jet structure~\cite{Soyez:2011np}.
Since the jet spectra in the numerator are always with a smaller jet resolution parameter $R$, it is expected that QCD radiation reduces this ratio below one, and that the effect decreases with the increasing collimation of jets at high \pt~\cite{CMS:2014nvq}.
The ratios confirm the expected trend of increasing collimation with increasing transverse momentum of jets, corroborated also by the theoretical predictions.
Figure~\ref{fig:xsecratioppVspPb} shows the comparison of the cross section ratios in \pp collisions and \pPb collisions.
The comparison shows that the energy in the jet cone distributed transverse to the jet axis in \pPb collisions is consistent with that in \pp collisions.
No sign of a modified jet structure is observed within uncertainties.

\subsection{Nuclear modification factor $\RpPbjch$}

The nuclear modification factor of charged-particle jets in minimum bias \pPb collisions due to nuclear matter effects is quantified by comparing the jet cross section in \pPb collisions normalized by the number of nucleons of the Pb ion, $A=208$, to the jet cross section in \pp collisions, known as the $A$-scaling hypothesis~\cite{CMS:2022wfi, Miller:2007ri},
\begin{equation}\label{eq:RpPb}
\RpPbjch = \left. \frac{1}{A} \frac{\dd^{2}\sigma_{\rm pPb}}{\dd\pTjch\dd\etaj} \right / \frac{\dd^{2}\sigma_{\rm pp}}{\dd\pTjch\dd\etaj}.
\end{equation}
The jet cross sections are measured in the laboratory frame with $\abs{\etaj} < 0.5$ in both \pp and \pPb collisions.
As mentioned above, the laboratory frame is shifted from the center-of-mass frame in rapidity by $\Delta y = 0.465$ in \pPb collisions while it is not shifted for \pp collisions, resulting in different jet rapidity acceptances in the center-of-mass frame between the two systems.
However, this effect on $\RpPbjch$ is smaller than $5\%$~\cite{ALICE:2015umm}, and it is not accounted in Eq.~(\ref{eq:RpPb}).

\begin{figure}
\centering
\includegraphics[width=\textwidth]{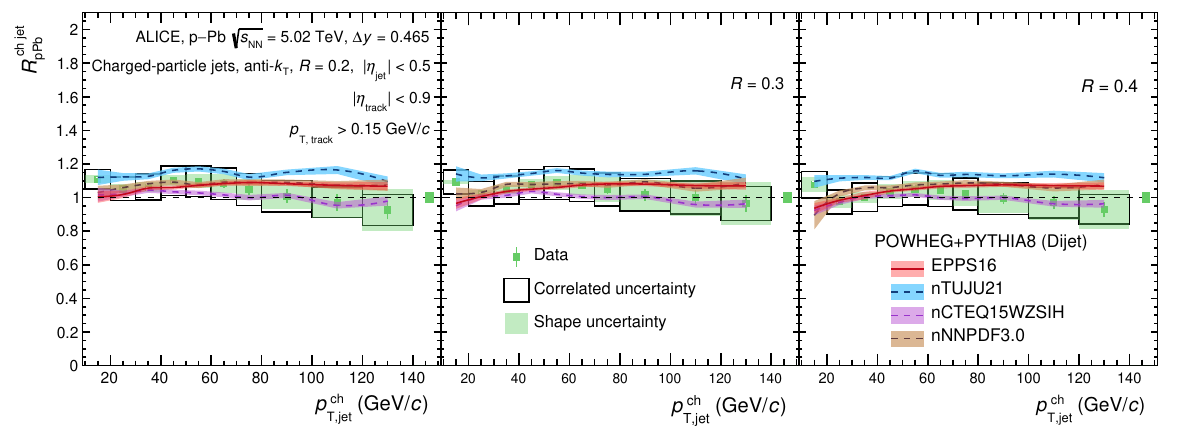}
\caption{The nuclear modification factor $\RpPbjch$ of inclusive charged-particle jets as a function of \pTjch at \fivenn for $R=0.2$, $0.3$, and $0.4$.
The data measured are compared with the \Pwgpy predictions calculated with various nPDFs. The bands on the predictions represent statistical uncertainties.
Systematic and statistical uncertainties of data are shown as boxes and error bars, respectively.
The normalization uncertainty of $4.37\%$ is shown as a box around one.}
\label{fig:RpPb}
\end{figure}

Figure~\ref{fig:RpPb} depicts the nuclear modification factors $\RpPbjch$ for jets with $R=0.2$, $0.3$, and $0.4$ as a function of jet transverse momentum.
The $\RpPbjch$ is compatible with one within uncertainties in the reported transverse momentum range $10<\pTjch<140$~\GeVc, and it is noted to be approximately independent of jet transverse momentum and jet resolution parameter.
The $\RpPbjch$ presented in this article is in agreement with the Run $1$ result~\cite{ALICE:2015umm} within uncertainties.
The results also indicate that jet quenching, if present, is below the sensitivity of the current measurement.
The \Pwgpy prediction calculated using the nCTEQ$15$WZSIH is found to best describe the data within uncertainties.
This shows that the effects of nuclear-modified PDFs introduced by the nPDFs have minor impact on jet production.
The data from this measurement can provide an important constraint on the global analysis of nPDFs.

\begin{figure}
\centering
\includegraphics[width=.7\textwidth]{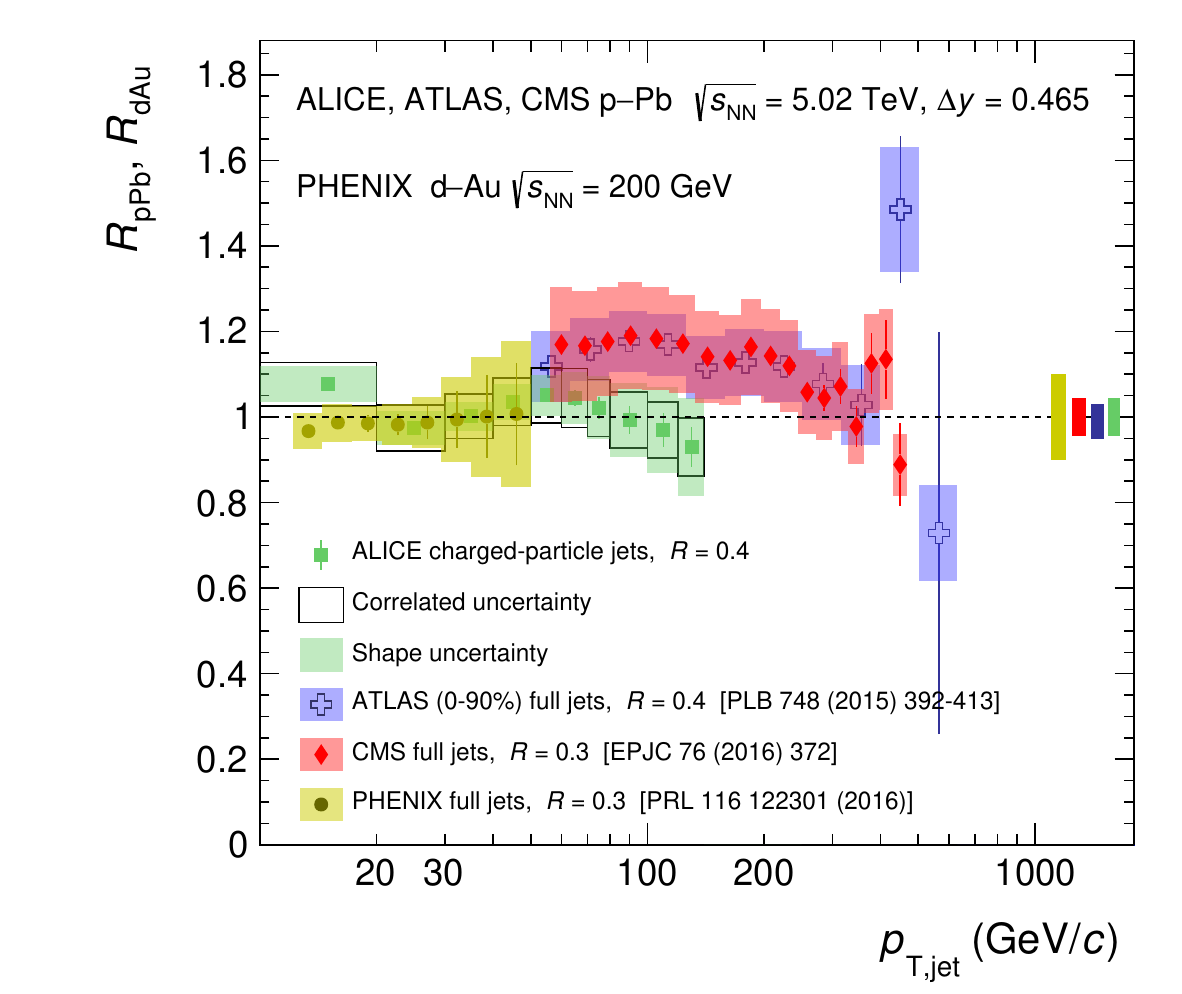}
\caption{Comparison of the nuclear modification factors of jets in \pPb and \dAu measurements at the LHC and RHIC, respectively.
The boxes around the data points denote the systematic uncertainties, while the error bars denote the statistical uncertainties.
The systematic uncertainties on the normalization are shown as boxes at $R_{\rm pPb}=1$.}
\label{fig:RpPbComparison}
\end{figure}

The $\RpPbjch$ result reported in this paper is compared to other published experimental results currently available.
Figure~\ref{fig:RpPbComparison} shows the comparison to the measurement of full jets in \pPb collisions at $\sNN = 5.02$~\TeV by the ATLAS~\cite{ATLAS:2014cpa} and CMS Collaborations~\cite{CMS:2016svx} at the LHC, and \dAu collisions at $\sNN = 200$~\GeV by the PHENIX Collaboration~\cite{PHENIX:2015fgy} at RHIC.
Full jets are reconstructed with charged and neutral components.
It is important to realize that the energy scales of the ATLAS, CMS, PHENIX, and ALICE measurements are different (jets measured by ALICE do not include neutral fragments) which complicates a direct comparison between the measurements.
The ATLAS and CMS measurements show a hint of enhancement above one, but it has to be confirmed with higher precision measurements.
It is worth noticing that by assuming the final state particles are dominated by pions, one can roughly estimate a scaling factor of around $1.5$ between the energy of charged-particle jets and that of full jets.
In general, the ALICE measurement is in qualitative agreement with those from ATLAS and CMS within the current experimental precision.
The ALICE results shown here extend the measurements down to a jet $\pt$ of $10$~\GeVc and complement the measurements of the ATLAS and CMS Collaborations.

\section{Conclusion}%
\label{sec:con}

The inclusive \pt-differential charged-particle jet production cross sections in \pp and \pPb collisions at \fivenn were measured using the ALICE detector at the LHC.
The inclusive charged-particle jets were reconstructed with resolution parameters $R=0.2$, $0.3$, and $0.4$.
The measured charged-particle jet cross sections are corrected for experimental effects, such as the finite detector resolution on the jet energy scale as well as the effects of the uncorrelated background and its fluctuations.
The ratios of jet cross sections measured for different values of $R$ in \pp collisions are consistent with those in \pPb collisions within uncertainties, indicating no sign of jet structure modification in \pPb collisions within the current measurement precision.
Besides, the results confirm that the higher-$\pt$ jets are more collimated.
The cross section ratios also provide additional comparisons to theoretical predictions.

Within the current experimental precision and uncertainties, the nuclear modification  factor $\RpPbjch$ is observed to be consistent with one, implying that the nuclear effects on jet production in \pPb collisions are below the resolution of the current measurement.
The $\RpPbjch$ is also found to be approximately independent of the jet resolution parameter, and is consistent with the measurements of full jets by the ATLAS, CMS, and PHENIX Collaborations within the kinematic region of overlap among the different measurements.
The ALICE results reported in this paper extend the jet $\pt$ reach down to $10$~\GeVc and are thus complementary to those obtained with ATLAS and CMS.

The results are well described by NLO \Pwgpy predictions (for \pp and \pPb collisions), while the \Scape (for \pp collisions) prediction agrees better with the data at high \pt.
These results provide a constraint on initial- and final-state effects in nuclear collisions, global analysis of nPDF, and provide a new baseline for the study of jet production in heavy-ion collisions.


\newenvironment{acknowledgement}{\relax}{\relax}
\begin{acknowledgement}
\section*{Acknowledgements}

The ALICE Collaboration would like to thank all its engineers and technicians for their invaluable contributions to the construction of the experiment and the CERN accelerator teams for the outstanding performance of the LHC complex.
The ALICE Collaboration gratefully acknowledges the resources and support provided by all Grid centres and the Worldwide LHC Computing Grid (WLCG) collaboration.
The ALICE Collaboration acknowledges the following funding agencies for their support in building and running the ALICE detector:
A. I. Alikhanyan National Science Laboratory (Yerevan Physics Institute) Foundation (ANSL), State Committee of Science and World Federation of Scientists (WFS), Armenia;
Austrian Academy of Sciences, Austrian Science Fund (FWF): [M 2467-N36] and Nationalstiftung f\"{u}r Forschung, Technologie und Entwicklung, Austria;
Ministry of Communications and High Technologies, National Nuclear Research Center, Azerbaijan;
Conselho Nacional de Desenvolvimento Cient\'{\i}fico e Tecnol\'{o}gico (CNPq), Financiadora de Estudos e Projetos (Finep), Funda\c{c}\~{a}o de Amparo \`{a} Pesquisa do Estado de S\~{a}o Paulo (FAPESP) and Universidade Federal do Rio Grande do Sul (UFRGS), Brazil;
Bulgarian Ministry of Education and Science, within the National Roadmap for Research Infrastructures 2020-2027 (object CERN), Bulgaria;
Ministry of Education of China (MOEC) , Ministry of Science \& Technology of China (MSTC) and National Natural Science Foundation of China (NSFC), China;
Ministry of Science and Education and Croatian Science Foundation, Croatia;
Centro de Aplicaciones Tecnol\'{o}gicas y Desarrollo Nuclear (CEADEN), Cubaenerg\'{\i}a, Cuba;
Ministry of Education, Youth and Sports of the Czech Republic, Czech Republic;
The Danish Council for Independent Research | Natural Sciences, the VILLUM FONDEN and Danish National Research Foundation (DNRF), Denmark;
Helsinki Institute of Physics (HIP), Finland;
Commissariat \`{a} l'Energie Atomique (CEA) and Institut National de Physique Nucl\'{e}aire et de Physique des Particules (IN2P3) and Centre National de la Recherche Scientifique (CNRS), France;
Bundesministerium f\"{u}r Bildung und Forschung (BMBF) and GSI Helmholtzzentrum f\"{u}r Schwerionenforschung GmbH, Germany;
General Secretariat for Research and Technology, Ministry of Education, Research and Religions, Greece;
National Research, Development and Innovation Office, Hungary;
Department of Atomic Energy Government of India (DAE), Department of Science and Technology, Government of India (DST), University Grants Commission, Government of India (UGC) and Council of Scientific and Industrial Research (CSIR), India;
National Research and Innovation Agency - BRIN, Indonesia;
Istituto Nazionale di Fisica Nucleare (INFN), Italy;
Japanese Ministry of Education, Culture, Sports, Science and Technology (MEXT) and Japan Society for the Promotion of Science (JSPS) KAKENHI, Japan;
Consejo Nacional de Ciencia (CONACYT) y Tecnolog\'{i}a, through Fondo de Cooperaci\'{o}n Internacional en Ciencia y Tecnolog\'{i}a (FONCICYT) and Direcci\'{o}n General de Asuntos del Personal Academico (DGAPA), Mexico;
Nederlandse Organisatie voor Wetenschappelijk Onderzoek (NWO), Netherlands;
The Research Council of Norway, Norway;
Commission on Science and Technology for Sustainable Development in the South (COMSATS), Pakistan;
Pontificia Universidad Cat\'{o}lica del Per\'{u}, Peru;
Ministry of Education and Science, National Science Centre and WUT ID-UB, Poland;
Korea Institute of Science and Technology Information and National Research Foundation of Korea (NRF), Republic of Korea;
Ministry of Education and Scientific Research, Institute of Atomic Physics, Ministry of Research and Innovation and Institute of Atomic Physics and University Politehnica of Bucharest, Romania;
Ministry of Education, Science, Research and Sport of the Slovak Republic, Slovakia;
National Research Foundation of South Africa, South Africa;
Swedish Research Council (VR) and Knut \& Alice Wallenberg Foundation (KAW), Sweden;
European Organization for Nuclear Research, Switzerland;
Suranaree University of Technology (SUT), National Science and Technology Development Agency (NSTDA), Thailand Science Research and Innovation (TSRI) and National Science, Research and Innovation Fund (NSRF), Thailand;
Turkish Energy, Nuclear and Mineral Research Agency (TENMAK), Turkey;
National Academy of  Sciences of Ukraine, Ukraine;
Science and Technology Facilities Council (STFC), United Kingdom;
National Science Foundation of the United States of America (NSF) and United States Department of Energy, Office of Nuclear Physics (DOE NP), United States of America.
In addition, individual groups or members have received support from:
European Research Council, Strong 2020 - Horizon 2020 (grant nos. 950692, 824093), European Union;
Academy of Finland (Center of Excellence in Quark Matter) (grant nos. 346327, 346328), Finland.

\end{acknowledgement}

\bibliographystyle{etc/utphys}
\bibliography{AliJetpPbRun2}

\newpage
\appendix

\section{The ALICE Collaboration}
\label{app:collab}
\begin{flushleft} 
\small

S.~Acharya\,\orcidlink{0000-0002-9213-5329}\,$^{\rm 128}$, 
D.~Adamov\'{a}\,\orcidlink{0000-0002-0504-7428}\,$^{\rm 87}$, 
A.~Adler$^{\rm 71}$, 
G.~Aglieri Rinella\,\orcidlink{0000-0002-9611-3696}\,$^{\rm 33}$, 
M.~Agnello\,\orcidlink{0000-0002-0760-5075}\,$^{\rm 30}$, 
N.~Agrawal\,\orcidlink{0000-0003-0348-9836}\,$^{\rm 52}$, 
Z.~Ahammed\,\orcidlink{0000-0001-5241-7412}\,$^{\rm 136}$, 
S.~Ahmad\,\orcidlink{0000-0003-0497-5705}\,$^{\rm 16}$, 
S.U.~Ahn\,\orcidlink{0000-0001-8847-489X}\,$^{\rm 72}$, 
I.~Ahuja\,\orcidlink{0000-0002-4417-1392}\,$^{\rm 38}$
% ...

\section*{Affiliation Notes}

$^{\rm I}$ Deceased\\
$^{\rm II}$ Also at: Max-Planck-Institut fur Physik, Munich, Germany\\
$^{\rm III}$ Also at: Italian National Agency for New Technologies, Energy and Sustainable Economic Development (ENEA), Bologna, Italy\\
$^{\rm IV}$ Also at: Dipartimento DET del Politecnico di Torino, Turin, Italy\\
$^{\rm V}$ Also at: Department of Applied Physics, Aligarh Muslim University, Aligarh, India\\
$^{\rm VI}$ Also at: Institute of Theoretical Physics, University of Wroclaw, Poland\\
$^{\rm VII}$ Also at: An institution covered by a cooperation agreement with CERN\\

\section*{Collaboration Institutes}

$^{1}$ A.I. Alikhanyan National Science Laboratory (Yerevan Physics Institute) Foundation, Yerevan, Armenia\\
$^{2}$ AGH University of Krakow, Cracow, Poland\\
$^{3}$ Bogolyubov Institute for Theoretical Physics, National Academy of Sciences of Ukraine, Kiev, Ukraine\\
$^{4}$ Bose Institute, Department of Physics  and Centre for Astroparticle Physics and Space Science (CAPSS), Kolkata, India\\
$^{5}$ California Polytechnic State University, San Luis Obispo, California, United States\\
$^{6}$ Central China Normal University, Wuhan, China\\
$^{7}$ Centro de Aplicaciones Tecnol\'{o}gicas y Desarrollo Nuclear (CEADEN), Havana, Cuba\\
$^{8}$ Centro de Investigaci\'{o}n y de Estudios Avanzados (CINVESTAV), Mexico City and M\'{e}rida, Mexico\\
$^{9}$ Chicago State University, Chicago, Illinois, United States\\
$^{10}$ China Institute of Atomic Energy, Beijing, China\\
$^{11}$ China University of Geosciences, Wuhan, China\\
$^{12}$ Chungbuk National University, Cheongju, Republic of Korea\\
$^{13}$ Comenius University Bratislava, Faculty of Mathematics, Physics and Informatics, Bratislava, Slovak Republic\\
$^{14}$ COMSATS University Islamabad, Islamabad, Pakistan\\
$^{15}$ Creighton University, Omaha, Nebraska, United States\\
$^{16}$ Department of Physics, Aligarh Muslim University, Aligarh, India\\
$^{17}$ Department of Physics, Pusan National University, Pusan, Republic of Korea\\
$^{18}$ Department of Physics, Sejong University, Seoul, Republic of Korea\\
$^{19}$ Department of Physics, University of California, Berkeley, California, United States\\
$^{20}$ Department of Physics, University of Oslo, Oslo, Norway\\
$^{21}$ Department of Physics and Technology, University of Bergen, Bergen, Norway\\
$^{22}$ Dipartimento di Fisica, Universit\`{a} di Pavia, Pavia, Italy\\
$^{23}$ Dipartimento di Fisica dell'Universit\`{a} and Sezione INFN, Cagliari, Italy\\
$^{24}$ Dipartimento di Fisica dell'Universit\`{a} and Sezione INFN, Trieste, Italy\\
$^{25}$ Dipartimento di Fisica dell'Universit\`{a} and Sezione INFN, Turin, Italy\\
$^{26}$ Dipartimento di Fisica e Astronomia dell'Universit\`{a} and Sezione INFN, Bologna, Italy\\
$^{27}$ Dipartimento di Fisica e Astronomia dell'Universit\`{a} and Sezione INFN, Catania, Italy\\
$^{28}$ Dipartimento di Fisica e Astronomia dell'Universit\`{a} and Sezione INFN, Padova, Italy\\
$^{29}$ Dipartimento di Fisica `E.R.~Caianiello' dell'Universit\`{a} and Gruppo Collegato INFN, Salerno, Italy\\
$^{30}$ Dipartimento DISAT del Politecnico and Sezione INFN, Turin, Italy\\
$^{31}$ Dipartimento di Scienze MIFT, Universit\`{a} di Messina, Messina, Italy\\
$^{32}$ Dipartimento Interateneo di Fisica `M.~Merlin' and Sezione INFN, Bari, Italy\\
$^{33}$ European Organization for Nuclear Research (CERN), Geneva, Switzerland\\
$^{34}$ Faculty of Electrical Engineering, Mechanical Engineering and Naval Architecture, University of Split, Split, Croatia\\
$^{35}$ Faculty of Engineering and Science, Western Norway University of Applied Sciences, Bergen, Norway\\
$^{36}$ Faculty of Nuclear Sciences and Physical Engineering, Czech Technical University in Prague, Prague, Czech Republic\\
$^{37}$ Faculty of Physics, Sofia University, Sofia, Bulgaria\\
$^{38}$ Faculty of Science, P.J.~\v{S}af\'{a}rik University, Ko\v{s}ice, Slovak Republic\\
$^{39}$ Frankfurt Institute for Advanced Studies, Johann Wolfgang Goethe-Universit\"{a}t Frankfurt, Frankfurt, Germany\\
$^{40}$ Fudan University, Shanghai, China\\
$^{41}$ Gangneung-Wonju National University, Gangneung, Republic of Korea\\
$^{42}$ Gauhati University, Department of Physics, Guwahati, India\\
$^{43}$ Helmholtz-Institut f\"{u}r Strahlen- und Kernphysik, Rheinische Friedrich-Wilhelms-Universit\"{a}t Bonn, Bonn, Germany\\
$^{44}$ Helsinki Institute of Physics (HIP), Helsinki, Finland\\
$^{45}$ High Energy Physics Group,  Universidad Aut\'{o}noma de Puebla, Puebla, Mexico\\
$^{46}$ Horia Hulubei National Institute of Physics and Nuclear Engineering, Bucharest, Romania\\
$^{47}$ HUN-REN Wigner Research Centre for Physics, Budapest, Hungary\\
$^{48}$ Indian Institute of Technology Bombay (IIT), Mumbai, India\\
$^{49}$ Indian Institute of Technology Indore, Indore, India\\
$^{50}$ INFN, Laboratori Nazionali di Frascati, Frascati, Italy\\
$^{51}$ INFN, Sezione di Bari, Bari, Italy\\
$^{52}$ INFN, Sezione di Bologna, Bologna, Italy\\
$^{53}$ INFN, Sezione di Cagliari, Cagliari, Italy\\
$^{54}$ INFN, Sezione di Catania, Catania, Italy\\
$^{55}$ INFN, Sezione di Padova, Padova, Italy\\
$^{56}$ INFN, Sezione di Pavia, Pavia, Italy\\
$^{57}$ INFN, Sezione di Torino, Turin, Italy\\
$^{58}$ INFN, Sezione di Trieste, Trieste, Italy\\
$^{59}$ Inha University, Incheon, Republic of Korea\\
$^{60}$ Institute for Gravitational and Subatomic Physics (GRASP), Utrecht University/Nikhef, Utrecht, Netherlands\\
$^{61}$ Institute of Experimental Physics, Slovak Academy of Sciences, Ko\v{s}ice, Slovak Republic\\
$^{62}$ Institute of Physics, Homi Bhabha National Institute, Bhubaneswar, India\\
$^{63}$ Institute of Physics of the Czech Academy of Sciences, Prague, Czech Republic\\
$^{64}$ Institute of Space Science (ISS), Bucharest, Romania\\
$^{65}$ Institut f\"{u}r Kernphysik, Johann Wolfgang Goethe-Universit\"{a}t Frankfurt, Frankfurt, Germany\\
$^{66}$ Instituto de Ciencias Nucleares, Universidad Nacional Aut\'{o}noma de M\'{e}xico, Mexico City, Mexico\\
$^{67}$ Instituto de F\'{i}sica, Universidade Federal do Rio Grande do Sul (UFRGS), Porto Alegre, Brazil\\
$^{68}$ Instituto de F\'{\i}sica, Universidad Nacional Aut\'{o}noma de M\'{e}xico, Mexico City, Mexico\\
$^{69}$ iThemba LABS, National Research Foundation, Somerset West, South Africa\\
$^{70}$ Jeonbuk National University, Jeonju, Republic of Korea\\
$^{71}$ Johann-Wolfgang-Goethe Universit\"{a}t Frankfurt Institut f\"{u}r Informatik, Fachbereich Informatik und Mathematik, Frankfurt, Germany\\
$^{72}$ Korea Institute of Science and Technology Information, Daejeon, Republic of Korea\\
$^{73}$ KTO Karatay University, Konya, Turkey\\
$^{74}$ Laboratoire de Physique Subatomique et de Cosmologie, Universit\'{e} Grenoble-Alpes, CNRS-IN2P3, Grenoble, France\\
$^{75}$ Lawrence Berkeley National Laboratory, Berkeley, California, United States\\
$^{76}$ Lund University Department of Physics, Division of Particle Physics, Lund, Sweden\\
$^{77}$ Nagasaki Institute of Applied Science, Nagasaki, Japan\\
$^{78}$ Nara Women{'}s University (NWU), Nara, Japan\\
$^{79}$ National and Kapodistrian University of Athens, School of Science, Department of Physics , Athens, Greece\\
$^{80}$ National Centre for Nuclear Research, Warsaw, Poland\\
$^{81}$ National Institute of Science Education and Research, Homi Bhabha National Institute, Jatni, India\\
$^{82}$ National Nuclear Research Center, Baku, Azerbaijan\\
$^{83}$ National Research and Innovation Agency - BRIN, Jakarta, Indonesia\\
$^{84}$ Niels Bohr Institute, University of Copenhagen, Copenhagen, Denmark\\
$^{85}$ Nikhef, National institute for subatomic physics, Amsterdam, Netherlands\\
$^{86}$ Nuclear Physics Group, STFC Daresbury Laboratory, Daresbury, United Kingdom\\
$^{87}$ Nuclear Physics Institute of the Czech Academy of Sciences, Husinec-\v{R}e\v{z}, Czech Republic\\
$^{88}$ Oak Ridge National Laboratory, Oak Ridge, Tennessee, United States\\
$^{89}$ Ohio State University, Columbus, Ohio, United States\\
$^{90}$ Physics department, Faculty of science, University of Zagreb, Zagreb, Croatia\\
$^{91}$ Physics Department, Panjab University, Chandigarh, India\\
$^{92}$ Physics Department, University of Jammu, Jammu, India\\
$^{93}$ Physics Program and International Institute for Sustainability with Knotted Chiral Meta Matter (SKCM2), Hiroshima University, Hiroshima, Japan\\
$^{94}$ Physikalisches Institut, Eberhard-Karls-Universit\"{a}t T\"{u}bingen, T\"{u}bingen, Germany\\
$^{95}$ Physikalisches Institut, Ruprecht-Karls-Universit\"{a}t Heidelberg, Heidelberg, Germany\\
$^{96}$ Physik Department, Technische Universit\"{a}t M\"{u}nchen, Munich, Germany\\
$^{97}$ Politecnico di Bari and Sezione INFN, Bari, Italy\\
$^{98}$ Research Division and ExtreMe Matter Institute EMMI, GSI Helmholtzzentrum f\"ur Schwerionenforschung GmbH, Darmstadt, Germany\\
$^{99}$ Saga University, Saga, Japan\\
$^{100}$ Saha Institute of Nuclear Physics, Homi Bhabha National Institute, Kolkata, India\\
$^{101}$ School of Physics and Astronomy, University of Birmingham, Birmingham, United Kingdom\\
$^{102}$ Secci\'{o}n F\'{\i}sica, Departamento de Ciencias, Pontificia Universidad Cat\'{o}lica del Per\'{u}, Lima, Peru\\
$^{103}$ Stefan Meyer Institut f\"{u}r Subatomare Physik (SMI), Vienna, Austria\\
$^{104}$ SUBATECH, IMT Atlantique, Nantes Universit\'{e}, CNRS-IN2P3, Nantes, France\\
$^{105}$ Sungkyunkwan University, Suwon City, Republic of Korea\\
$^{106}$ Suranaree University of Technology, Nakhon Ratchasima, Thailand\\
$^{107}$ Technical University of Ko\v{s}ice, Ko\v{s}ice, Slovak Republic\\
$^{108}$ The Henryk Niewodniczanski Institute of Nuclear Physics, Polish Academy of Sciences, Cracow, Poland\\
$^{109}$ The University of Texas at Austin, Austin, Texas, United States\\
$^{110}$ Universidad Aut\'{o}noma de Sinaloa, Culiac\'{a}n, Mexico\\
$^{111}$ Universidade de S\~{a}o Paulo (USP), S\~{a}o Paulo, Brazil\\
$^{112}$ Universidade Estadual de Campinas (UNICAMP), Campinas, Brazil\\
$^{113}$ Universidade Federal do ABC, Santo Andre, Brazil\\
$^{114}$ Universitatea Nationala de Stiinta si Tehnologie Politehnica Bucuresti, Bucharest, Romania\\
$^{115}$ University of Cape Town, Cape Town, South Africa\\
$^{116}$ University of Derby, Derby, United Kingdom\\
$^{117}$ University of Houston, Houston, Texas, United States\\
$^{118}$ University of Jyv\"{a}skyl\"{a}, Jyv\"{a}skyl\"{a}, Finland\\
$^{119}$ University of Kansas, Lawrence, Kansas, United States\\
$^{120}$ University of Liverpool, Liverpool, United Kingdom\\
$^{121}$ University of Science and Technology of China, Hefei, China\\
$^{122}$ University of South-Eastern Norway, Kongsberg, Norway\\
$^{123}$ University of Tennessee, Knoxville, Tennessee, United States\\
$^{124}$ University of the Witwatersrand, Johannesburg, South Africa\\
$^{125}$ University of Tokyo, Tokyo, Japan\\
$^{126}$ University of Tsukuba, Tsukuba, Japan\\
$^{127}$ Universit\"{a}t M\"{u}nster, Institut f\"{u}r Kernphysik, M\"{u}nster, Germany\\
$^{128}$ Universit\'{e} Clermont Auvergne, CNRS/IN2P3, LPC, Clermont-Ferrand, France\\
$^{129}$ Universit\'{e} de Lyon, CNRS/IN2P3, Institut de Physique des 2 Infinis de Lyon, Lyon, France\\
$^{130}$ Universit\'{e} de Strasbourg, CNRS, IPHC UMR 7178, F-67000 Strasbourg, France, Strasbourg, France\\
$^{131}$ Universit\'{e} Paris-Saclay, Centre d'Etudes de Saclay (CEA), IRFU, D\'{e}partment de Physique Nucl\'{e}aire (DPhN), Saclay, France\\
$^{132}$ Universit\'{e}  Paris-Saclay, CNRS/IN2P3, IJCLab, Orsay, France\\
$^{133}$ Universit\`{a} degli Studi di Foggia, Foggia, Italy\\
$^{134}$ Universit\`{a} del Piemonte Orientale, Vercelli, Italy\\
$^{135}$ Universit\`{a} di Brescia, Brescia, Italy\\
$^{136}$ Variable Energy Cyclotron Centre, Homi Bhabha National Institute, Kolkata, India\\
$^{137}$ Warsaw University of Technology, Warsaw, Poland\\
$^{138}$ Wayne State University, Detroit, Michigan, United States\\
$^{139}$ Yale University, New Haven, Connecticut, United States\\
$^{140}$ Yonsei University, Seoul, Republic of Korea\\
$^{141}$  Zentrum  f\"{u}r Technologie und Transfer (ZTT), Worms, Germany\\
$^{142}$ Affiliated with an institute covered by a cooperation agreement with CERN\\
$^{143}$ Affiliated with an international laboratory covered by a cooperation agreement with CERN.\\

\end{flushleft} 

\end{document}